\documentclass[letterpaper]{article}
\usepackage[preprint]{aaai2027}
\usepackage[hyphens]{url}
\usepackage{graphicx}
\usepackage{natbib}
\usepackage{caption}
\usepackage{booktabs}
\usepackage{amsmath}
\usepackage{amssymb}
\usepackage{multirow}
\usepackage{placeins}

\newcommand{\method}{FinDeCrowd-RAG}
\newcommand{\stress}{FinDeCrowd-Stress}
\newcommand{\matcheddrop}{0.147}
\newcommand{\controlledgain}{0.145}
\newcommand{\naturalbase}{0.420}
\newcommand{\naturalgated}{0.492}

\title{Financial Evidence Crowding:\\
Diagnosing and Mitigating Constraint-Induced Displacement in Retrieval-Augmented Generation}
\author{
Yixi Zhou\textsuperscript{\rm 1}\equalcontrib,
Jiayi Yin\textsuperscript{\rm 2}\equalcontrib,
Fan Zhang\textsuperscript{\rm 3},
Xiangyi He\textsuperscript{\rm 1},
Haipeng Zhang\textsuperscript{\rm 1}\corresponding
}
\affiliations{
\textsuperscript{\rm 1}ShanghaiTech University \quad
\textsuperscript{\rm 2}University of Wisconsin--Madison \quad
\textsuperscript{\rm 3}The University of Tokyo\\
\{zhouyx2022,hexy2025,zhanghp\}@shanghaitech.edu.cn, jyin66@wisc.edu,\\
zhang-fan@g.ecc.u-tokyo.ac.jp
}

\begin{document}
\maketitle
\begin{abstract}
Retrieval-augmented generation (RAG) retrieves candidate evidence and sends only a limited top-ranked subset, the top-$k$ context, to a generator.
In financial question answering, passages can match a query's topic while conflicting with its period, segment, metric scope, or table scope.
We study the resulting set-level ordering failure, which we call \emph{financial evidence crowding}.
\stress{} isolates this failure through matched compatible and incompatible candidate pools while fixing the query, relevant evidence, ranking model, candidate count, and retrieval budget.
On a FinDER test split containing companies unseen during training, incompatible pools reduce top-10 evidence inclusion (Recall@10) by \matcheddrop{} relative to equally difficult compatible pools.
This gap shows that conflicting candidates consume limited context slots and displace answer-supporting evidence.
We then introduce \method{}, a learned score correction that combines a fixed relevance score with typed compatibility and local lexical competition.
A query-level identity gate applies the correction only when it predicts a better order; otherwise, it preserves the original ranking.
On identical controlled candidates, \method{} raises top-10 evidence inclusion from 0.757 to 0.902 by recovering evidence already present in the candidate set.
On a FinDER index built without query-specific candidate insertion, gated reranking raises this inclusion rate from \naturalbase{} to \naturalgated{}, while top-100 retrieval coverage remains 0.743 by design.
With a fixed generator, the same ordering change improves answer accuracy and citation recall on FinanceBench and FinQA.
These results identify constraint-induced displacement as a measurable RAG evaluation target and show that identity-gated reranking can recover evidence already covered by first-stage retrieval.
\end{abstract}

\begin{figure}[!t]
\centering
\includegraphics[width=\columnwidth]{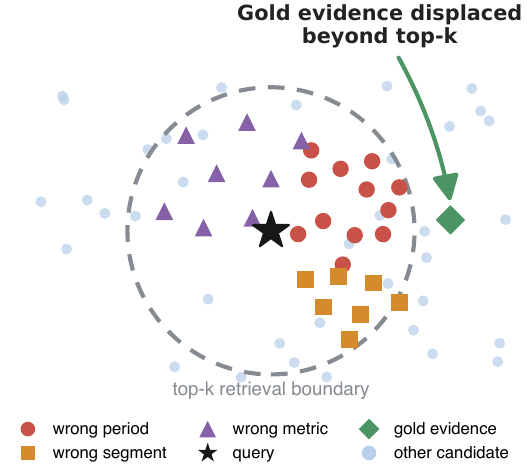}
\caption{Financial evidence crowding. Topically similar candidates with an observed constraint conflict displace answer-bearing evidence beyond the fixed top-$k$ boundary.}
\label{fig:financial-crowding}
\end{figure}

\begin{figure*}[t]
\centering
\includegraphics[width=\textwidth]{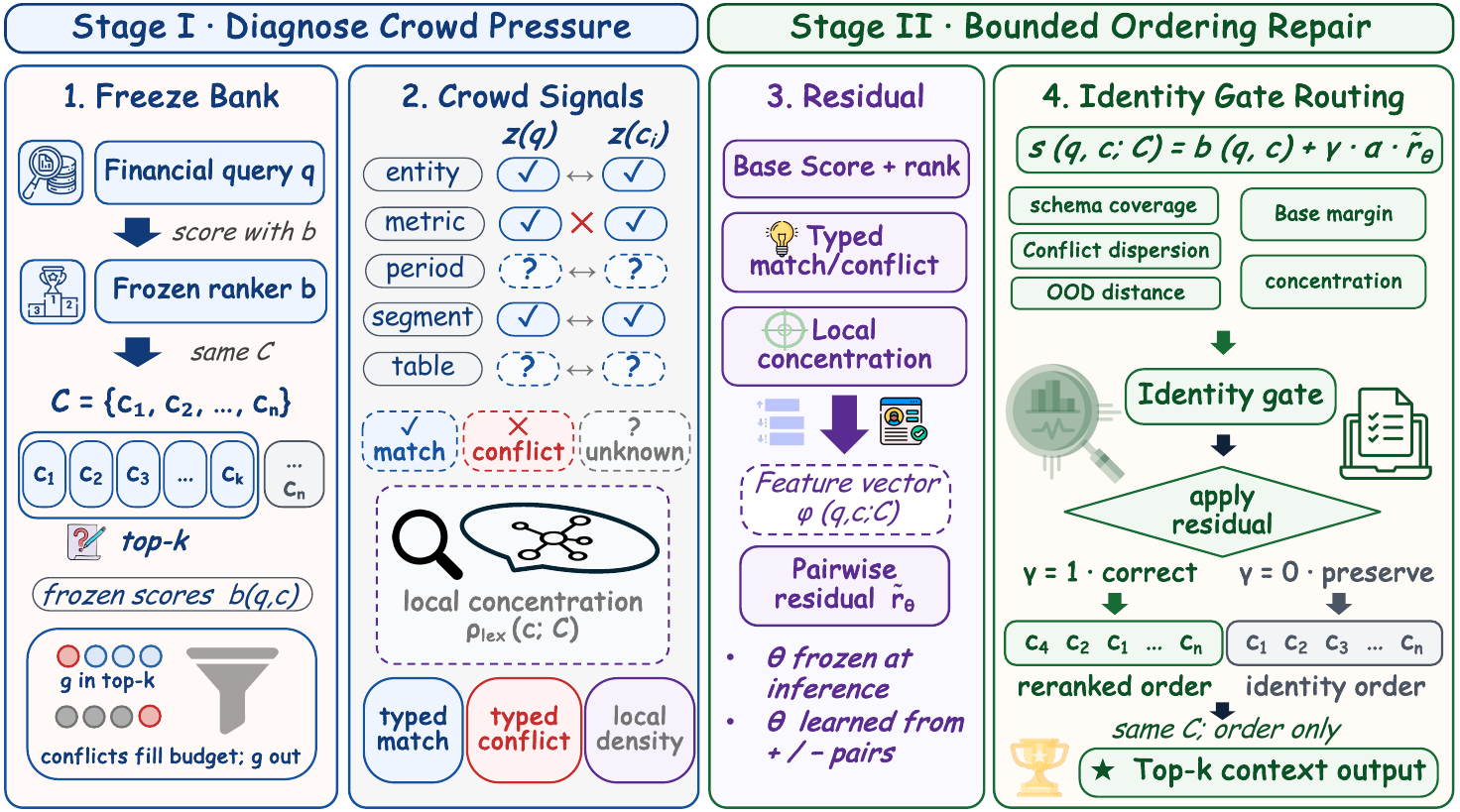}
\caption{\textsc{FinDeCrowd} first diagnoses displacement with matched compatible and incompatible pools, then reorders the same candidate bank with a learned score correction. The identity gate preserves the original order when the correction lacks support.}
\label{fig:overview}
\end{figure*}

\section{Introduction}

Financial questions often impose several evidence constraints.
Supporting evidence must match applicable query constraints on entity, metric, segment, period, and table scope.
Retrieval-augmented generation (RAG) retrieves a candidate bank, ranks its passages, and sends only the top $k$ to the generator.
A fixed context budget limits passage count; ranking selects what the generator can use.
A topical match may share the company and metric but report a different period or segment.
Conflicting passages in the top-$k$ slots can push supporting evidence beyond the generator's context.
Figure~\ref{fig:financial-crowding} shows this displacement.

Recent work shows how crowded embedding spaces marginalize RAG evidence~\citep{ashrafi2026the}.
Financial benchmarks report aggregate retrieval scores~\citep{islam2023financebench,choi2025finder,choe2025hirec,strich2025t2ragbench}.
These scores conflate candidate-set size and semantic competition with explicit query-constraint conflict.
These causes require different remedies.
Financial evidence makes the distinction observable because period, segment, metric, and table constraints often appear in queries and candidates.

We call displacement caused by explicit constraint conflicts \emph{financial evidence crowding}.
Figure~\ref{fig:overview} maps diagnosis to correction.
\stress{} compares same-size compatible and incompatible pools at matched semantic difficulty.
\method{} reranks the fixed candidate bank without retrieving new passages.
An identity gate applies a learned score adjustment only when it predicts a better order.
Controlled diagnosis is the central contribution; the correction tests whether it repairs the loss.

This paper makes three contributions:
\begin{itemize}
    \item \textbf{Controlled diagnosis.} We define a ranker-, budget-, and intervention-specific crowding contrast; with 30 distractors, incompatible pools lower Recall@10 by \matcheddrop{} and push the highest-ranked supporting unit 3.34 positions lower.
    \item \textbf{Fixed-bank correction.} A residual learned from supporting and distracting candidates raises controlled Recall@10 by \controlledgain{} on identical candidates and exceeds in-domain supervised, pretrained, large language model (LLM), and schema-aware rerankers.
    \item \textbf{Natural-bank and downstream evidence.} In a non-injected FinDER index, crowding accounts for 43.5\% of top-10 failures whose evidence remains in the top 100 (orderable failures); gated correction preserves coverage and improves ordering and answer support.
\end{itemize}

\section{Related Work}

\paragraph{Financial evidence retrieval.}
FinQA, TAT-QA, ConvFinQA, and MultiHiertt study reasoning over financial text and tables
~\citep{chen-etal-2021-finqa,zhu-etal-2021-tat,chen-etal-2022-convfinqa,zhao-etal-2022-multihiertt}.
FinanceBench, FinDER, LOFin, and T$^2$-RAGBench make evidence retrieval explicit
~\citep{islam2023financebench,choi2025finder,choe2025hirec,strich2025t2ragbench}.
These benchmark-level aggregate scores do not identify explicit constraint conflict separately from semantic competition.
\stress{} supplies that missing controlled contrast.

\paragraph{Hard negatives and neighborhood competition.}
Hard negatives are irrelevant passages selected because they resemble relevant evidence.
BEIR evaluates retrieval generalization, ANCE and RocketQA train with such difficult negatives, and ColBERTv2 uses token-level query--passage matching
~\citep{thakur2021beir,xiong2021ance,santhanam-etal-2022-colbertv2,qu-etal-2021-rocketqa}.
RARE measures retrieval under redundant evidence.
Hubness, where a few vector representations become frequent neighbors, and crowded-space analyses characterize dense $k$-nearest-neighbor neighborhoods
~\citep{cho-lee-2026-rare,nielsen2023hubness,ashrafi2026the}.
Long-context and compression studies show that evidence position and selection affect generation~\citep{liu-etal-2024-lost,xu2024recomp}.
Financial evidence crowding instead compares matched candidate sets that differ in observable compatibility, isolating top-$k$ allocation under a fixed ranker and budget.

\paragraph{Structure-aware reranking.}
Structure-aware and metadata driven retrieval uses entities, periods, table headers, and other fields
~\citep{li-etal-2023-structure,zhang-etal-2025-mrag,dadopoulos2025metadata}.
FinCARDS applies schema-aligned tournament reranking to financial evidence~\citep{zhou-etal-2026-fincards}.
These methods optimize overall retrieval quality.
\method{} serves a narrower role: it tests whether fixed-bank correction repairs a diagnosed, constraint-induced loss while preserving the candidate bank and base score under controlled evaluation settings.

\section{Financial Evidence Crowding: A Set-Level Ordering Failure}

Let $q$ denote a query, $C$ its candidate bank, $G_q\subseteq C$ its answer-supporting evidence, $b$ a scoring model that orders candidates, and $k$ the number sent to the generator.
A candidate is \emph{compatible} when its observed entity, metric, period, segment, table scope, unit, and currency fields contain no explicit conflict with $q$.
Compatibility permits irrelevant candidates; it only rules out observed contradiction.
We treat missing fields as unknown.

For a higher-is-better ordering metric $M_k$, \stress{} compares an incompatible pool with a same-size compatible pool:
\begin{equation}
\Delta_k(q;b,\mathcal I)
=M_k(b;q,C_{\mathrm{inc}})-M_k(b;q,C_{\mathrm{comp}}).
\label{eq:estimand}
\end{equation}
The intervention design $\mathcal I$ fixes the query, relevant evidence, ranker, candidate count, construction protocol, and matching population.
A negative $\Delta_k$ records additional displacement under incompatibility.
The estimand remains specific to $b$, $k$, and $\mathcal I$.

\stress{} builds typed interventions from FinDER evidence~\citep{choi2025finder}.
It extracts an intent schema, a normalized record of the query's financial constraints, and uses a designated gold unit only to fill fields missing during construction.
Metric, period, segment, and table-scope candidates match the entity, conflict on one target field, and preserve every other jointly observed field.
Compatible hard candidates match the observed constraints without supporting the answer.
\textsc{LexicalHard} selects hard negatives using term frequency--inverse document frequency (TF--IDF), a term-overlap representation that requires no financial-field parser.
We remove same-query evidence, duplicates, and near duplicates, then freeze a similarity-ordered candidate list before any evaluated ranker scores it.
Prefixes of this list create nested pools with 10, 30, and 60 distractors.

We use BGE reranker-base, a pretrained cross-encoder that reads a query--candidate pair and assigns a relevance score.
\emph{Frozen BGE} keeps this model's weights fixed; subsequent corrections change only the candidate order.
The primary comparison matches compatible and incompatible candidates on Frozen BGE similarity, TF--IDF similarity, company, table form, length, and numeric density.
Entity-disjoint splits keep each company in only one of train, development, and test; these splits precede construction and fitting.
All methods receive identical candidate IDs and relevance sets.
Three finance-trained annotators independently audit 600 shuffled items without ranks, pool names, or gold identifiers.
The evaluation reports both broad compatibility and exact conflict-type judgments.
The supplement provides eligibility, balance, gold-fill, matching, and annotation details.

For a non-injected bank, we use a separate operational label.
An \emph{orderable failure} places relevant evidence within ranks 11 through 100.
We label it crowded at $k=10$ only when at least one explicitly conflicting candidate ranks above the best relevant unit and deleting all such candidates, without rescoring, moves a relevant unit into the top 10.
This deletion test prevents us from labeling every covered top-10 miss as crowding.

\section{\method{}: Identity-Gated Constraint-Aware Correction}

\method{} tests whether fixed-bank reranking repairs the diagnosed allocation failure without retrieving new evidence.
It preserves the frozen relevance score and learns a \emph{residual} that adjusts the score before reranking candidates.
The residual combines the base score with lexical relevance, typed compatibility, schema coverage, and local lexical concentration.
A deterministic extractor maps queries and candidates to financial fields; its 480-record audit yields an average field-level F1 of 0.959, summarizing extraction precision and recall.
Missing observations contribute zero match and zero conflict.
The Supplement reports per-field scores and natural-bank slices.

Local lexical concentration measures how many textually similar competitors surround a candidate in the same bank.
For TF--IDF representation $v(c)$, local lexical concentration averages similarity to the five nearest candidates inside the evaluated bank:
\begin{equation}
\rho_{\mathrm{lex}}(c;C)=\frac{1}{5}
\sum_{c'\in\mathrm{NN}^{\mathrm{tfidf}}_5(c;C)}
\cos\!\left(v(c),v(c')\right).
\label{eq:concentration}
\end{equation}
Larger values indicate more repeated local competition for the same top-$k$ slots.
Crowding diagnosis uses only explicit conflicts; local concentration enters only the correction.

The final score adds a standardized linear residual to a within-bank normalized base score:
\begin{equation}
s(q,c;C)=\widetilde b(q,c)+\gamma(q,C)\alpha\widetilde r_\theta(q,c;C).
\label{eq:score}
\end{equation}
Here tildes denote within-bank standardization, $\alpha$ controls the correction scale, and $\gamma\in\{0,1\}$ controls activation.
Controlled experiments set $\gamma=1$.
We learn $\theta$ from pairs of supporting and distracting candidates, asking the residual to rank the support above at most 30 high-ranked distractors per query--pool bank.
Each query-pool contributes unit total weight.
Training uses Metric, Period, Segment, Table, and \textsc{LexicalHard} pools at $N=30$.
Training and inference use only query-candidate signals.

Always-on correction can improve crowded cases while disturbing already useful rankings.
The \emph{identity gate} is a binary router that either applies the residual or keeps the original order.
It predicts whether the residual improves normalized discounted cumulative gain at 10 (nDCG@10), a ranking score that rewards placing supporting evidence near the top.
Its inputs summarize available financial fields, detected conflicts, distance from training ranges, the top-score margin, and local concentration.
The gate label equals one when the residual strictly improves nDCG@10; ties receive zero.
FinDER calibrates the gate on 50 disjoint development queries.
Each new target domain requires a disjoint relevance-labeled calibration split, so the gate makes no zero-shot transfer claim.
The Supplement provides the signal ledger, fitting search, and gate protocol.
The gate makes one query-level decision.
The activated branch returns the residual order; the identity branch reproduces the base order exactly.
Both branches preserve candidate membership, so metric changes reflect ordering within a fixed candidate bank.

\section{Experiments}

\subsection{Evaluation Protocol}

\paragraph{Evaluation settings.}
We evaluate three claims under four settings.
Matched FinDER pools test whether incompatibility adds displacement.
Controlled FinDER pools test repair capacity on identical candidates.
Pools constructed without financial-field parsing from FinanceBench, FinQA, TAT-QA, and LOFin test transfer.
A non-injected FinDER global index measures occurrence and repair without query-specific evidence insertion.
The non-injected pipeline combines embedding-based dense retrieval with term-matching lexical retrieval, applies Frozen BGE, and retains 100 candidates.
As a second-stage utility test, a fixed generator consumes gated or identity-ordered evidence on disjoint FinanceBench and FinQA test subsets.

\paragraph{Baseline families.}
Pretrained neural baselines include Frozen BGE and Qwen3-Reranker-4B, a larger pretrained reranker~\citep{xiao2023cpack,chen2024bgem3,zhang2025qwen3embedding}.

The in-domain supervised baselines are LambdaMART, a tree-based learning-to-rank model, and fine-tuned BGE, a cross-encoder trained on the same FinDER pairs as \method{}.
LambdaMART and fine-tuned BGE use the same training pairs, entity-disjoint split, development labels, and selection metric as \method{}.

LLM-based baselines include REARANK-7B, a reasoning reranker, and RankGPT, which asks a language model to order a candidate list~\citep{zhang-etal-2025-rearank,sun-etal-2023-chatgpt}.

Finance-aware baselines include FinCARDS, a schema-guided tournament reranker, and LLM schema+BGE, which checks financial-field compatibility before applying BGE scores~\citep{zhou-etal-2026-fincards}.
Together, these categories test whether gains come from in-domain supervision, a larger pretrained reranker, LLM reasoning, or financial-schema use.
Every method receives the same test candidates, serialization, relevance labels, and cutoff.
The remaining baselines receive no additional FinDER supervision.
Test labels do not select models or thresholds.

\paragraph{Coverage and allocation metrics.}
Recall@100 measures retrieval coverage: whether the retrieved bank contains any supporting evidence among its first 100 candidates.
Recall@10 measures context allocation: whether supporting evidence reaches the ten passages available to the generator.
Because \method{} operates on a fixed bank, Recall@100 remains unchanged by design; Recall@10 gains isolate better allocation of already retrieved evidence.

\paragraph{Ordering and error metrics.}
nDCG@10 measures ordering quality within that context by giving more credit when supporting evidence appears earlier.
Best relevant rank records the position of the highest-ranked supporting unit, so a larger value indicates stronger displacement.
Repair rate is the share of crowded failures that gain top-10 evidence; regression rate is the share of queries whose nDCG@10 decreases after correction.
We average each controlled metric within a pool and then equally across pools.
Entity-disjoint splits, fixed candidate banks, and query-level inference preserve comparison parity.
The downstream study fixes Qwen2.5-7B-Instruct, an instruction-following language model, and changes only evidence order.
The Supplement specifies candidate construction, uncertainty, source separation, scoring, and annotation.

\subsection{Incompatibility Adds Displacement Beyond Hardness}

\paragraph{RQ1: Does financial constraint conflict cause displacement beyond semantic difficulty?}
Topically similar hard negatives and explicitly conflicting evidence can both occupy context slots.
We separate them by comparing same-size compatible and incompatible pools matched on semantic similarity and surface form.
The reported $\Delta$Recall@10 is incompatible minus compatible top-10 inclusion, so a negative value isolates additional displacement caused by conflict.

\begin{table}[t]
\centering
\footnotesize
\setlength{\tabcolsep}{2.6pt}
\begin{tabular}{@{}lrrl@{}}
\toprule
\multicolumn{4}{c}{\textit{Panel A: Construction checks}} \\
\midrule
Specification at $N=30$ & $Q$ & $\Delta$R@10 & 95\% CI \\
\midrule
Primary match & 400 & $-0.147$ & [$-0.164,-0.129$] \\
Zero-fill subset & 246 & $-0.142$ & [$-0.163,-0.121$] \\
Query-explicit construction & 265 & $-0.137$ & [$-0.157,-0.117$] \\
Exact target-observability & 251 & $-0.136$ & [$-0.157,-0.115$] \\
\midrule
\multicolumn{4}{c}{\textit{Panel B: Common-support checks}} \\
\midrule
Tight calipers & 312 & $-0.151$ & [$-0.173,-0.129$] \\
Wide calipers & 427 & $-0.139$ & [$-0.156,-0.122$] \\
Overlap weighting & 438 & $-0.144$ & [$-0.161,-0.127$] \\
Shuffled-label placebo & 400 & $-0.004$ & [$-0.017,0.010$] \\
\bottomrule
\end{tabular}
\caption{Matched-pool diagnostic. $Q$ is the evaluated query count, and CI is the query-level 95\% confidence interval. Panel A removes reliance on gold-filled fields. Panel B restricts or reweights comparisons to regions of similar difficulty; tight and wide calipers vary the allowed matching distance. Every non-placebo paired test has Holm-adjusted $p<0.001$.}
\label{tab:matched}
\end{table}

Table~\ref{tab:matched} shows that explicit conflict lowers top-10 evidence inclusion after matching semantic hardness.
The primary match balances Frozen BGE similarity (0.612 versus 0.611), TF--IDF similarity (0.427 versus 0.426), length, numeric density, company, and table form.
Compatible pools reach 0.883 Recall@10, while matched incompatible pools reach 0.736.
The \matcheddrop{} gap means that conflict lowers the top-10 evidence-inclusion rate by 14.7 percentage points.
The best supporting unit also moves 3.34 positions lower on average, confirming direct rank displacement.

Gold-free construction, target-observability matching, and three common-support estimators retain gaps between 0.136 and 0.151 (Table~\ref{tab:matched}).
The shuffled-label placebo centers near zero.
The field extractor reaches an average field-level F1 of 0.959, and the blinded construct audit reaches 96.5\% agreement; the Supplement reports the full audits.
These checks link the gap to compatibility after accounting for construction artifacts and unmatched difficulty.

Larger candidate pools intensify the same allocation failure.
From 10 to 60 distractors, Frozen BGE Recall@10 falls from 0.977 to 0.712 under structured conflicts and from 0.888 to 0.490 under lexical competition.
Dense-neighbor concentration, the degree to which candidates cluster in the base model's embedding space, also correlates with rank degradation at $r=0.347$; the Supplement reports the controlled regression.
Together, these results tie constraint-induced displacement to competition for a fixed top-10 budget.

\subsection{Fixed-Bank Correction Recovers Covered Evidence}

\paragraph{RQ2: Can ordering correction recover evidence without changing retrieval coverage?}
Every method receives the same fixed candidate bank, so none can add supporting passages.
Recall@10 gains therefore measure recovered context allocation, while nDCG@10 gains show whether supporting evidence also moves toward earlier positions.
The two ablations isolate schema and local-competition signals; the full residual combines them.

\begin{table}[t]
\centering
\footnotesize
\setlength{\tabcolsep}{1.7pt}
\begin{tabular}{@{}lrrr@{}}
\toprule
Method & R@10 & nDCG@10 & \textsc{LexHard} R@10 \\
\midrule
Frozen BGE & 0.757 & 0.557 & 0.576 \\
LambdaMART & 0.831 & 0.626 & 0.640 \\
Fine-tuned BGE & 0.849 & 0.651 & 0.657 \\
Qwen3-Reranker-4B & 0.842 & 0.648 & 0.653 \\
REARANK-7B & 0.847 & 0.657 & 0.658 \\
RankGPT & 0.836 & 0.645 & 0.646 \\
FinCARDS & 0.824 & 0.632 & 0.631 \\
LLM schema + BGE & 0.856 & 0.669 & 0.660 \\
\midrule
Conc.-related + base & 0.825 & 0.603 & 0.651 \\
Schema + base & 0.883 & 0.685 & 0.641 \\
\textbf{\method{}} & \textbf{0.902} & \textbf{0.714} & \textbf{0.681} \\
\bottomrule
\end{tabular}
\caption{Fixed-bank correction on identical candidates. R@10 measures top-10 inclusion, nDCG@10 rewards earlier supporting evidence, and \textsc{LexHard} tests parser-independent lexical competition.}
\label{tab:controlled}
\end{table}

We next compare stronger baselines and two simple rules: conflict-count ordering and a development-selected penalty.
Paired differences compare the same query banks, while query-level wins, ties, and losses reveal whether an average gain comes from broad improvement or a few large changes.

\begin{table}[t]
\centering
\footnotesize
\setlength{\tabcolsep}{1.0pt}
\begin{tabular}{@{}lrrr@{}}
\toprule
\multicolumn{4}{c}{\textit{Panel A: Paired strongest-baseline differences}} \\
\midrule
Control & $\Delta$R@10 & $\Delta$nDCG & nDCG W/T/L \\
\midrule
Fine-tuned BGE & 0.053 & 0.063 & 151/396/24 \\
REARANK-7B & 0.055 & 0.057 & 158/392/21 \\
LLM schema+BGE & 0.046 & 0.045 & 143/409/19 \\
\midrule
\multicolumn{4}{c}{\textit{Panel B: Simpler correction rules}} \\
\midrule
Method & R@10 & nDCG@10 & \textsc{LexHard} R@10 \\
\midrule
Conflict-count ordering & 0.866 & 0.668 & 0.633 \\
Dev-selected penalty & 0.879 & 0.684 & 0.650 \\
Full learned residual & 0.902 & 0.714 & 0.681 \\
\bottomrule
\end{tabular}
\caption{Direct tests of the learned correction. W/T/L counts query-level nDCG@10 wins, ties, and losses. Every Panel A difference has a 95\% CI above zero and Holm-adjusted $p<0.001$.}
\label{tab:controlled-depth}
\end{table}

\method{} raises Recall@10 from 0.757 to 0.902 on identical candidates (Table~\ref{tab:controlled}).
Because every method sees the same candidate bank, this 0.145 gain comes from restoring already retrieved evidence to the context window.
nDCG@10 rises from 0.557 to 0.714, showing that the recovered evidence also moves toward earlier positions.
The paired Recall@10 advantage over the three strongest controls ranges from 0.046 to 0.055 (Table~\ref{tab:controlled-depth}, Panel A).

Schema + base recovers most of the full-model gain in typed pools.
The concentration-related + base subset reaches 0.651 Recall@10 on \textsc{LexicalHard}, compared with 0.641 for Schema + base.
Simple conflict-count and fixed-penalty rules improve Frozen BGE but remain below the learned residual (Panel B).
Compatibility supplies most of the gain in typed pools, while concentration signals help most under lexical competition.
Their learned combination exceeds supervised, pretrained, LLM-based, and hand-specified alternatives on the same candidate banks.

\subsection{Correction Transfers Across Bases and Datasets}

\paragraph{Generality check: Does the diagnosed correction depend on one dataset or ranker?}
We test external financial datasets with parser-independent hard negatives and fit the same residual signal definition over four base rankers.
The external differences measure how much hard competition hurts top-10 inclusion and how much fixed-bank correction recovers.

\begin{table}[t]
\centering
\footnotesize
\setlength{\tabcolsep}{2.2pt}
\begin{tabular}{@{}lrrr@{}}
\toprule
\multicolumn{4}{c}{\textit{Panel A: External datasets}} \\
\midrule
Dataset & $Q$ & Hard$-$Random & Full$-$Hard \\
\midrule
FinanceBench & 150 & $-0.039$ & $+0.038$ \\
FinQA & 1,147 & $-0.126$ & $+0.046$ \\
TAT-QA & 1,660 & $-0.123$ & $+0.052$ \\
LOFin & 1,253 & $-0.104$ & $+0.035$ \\
\midrule
\multicolumn{4}{c}{\textit{Panel B: Base rankers}} \\
\midrule
Base ranker & Base & Full & $\Delta$R@10 \\
\midrule
Frozen BGE & 0.757 & 0.902 & +0.145 \\
Fine-tuned BGE & 0.849 & 0.914 & +0.065 \\
Qwen3-Reranker-4B & 0.842 & 0.909 & +0.067 \\
REARANK-7B & 0.847 & 0.911 & +0.064 \\
\bottomrule
\end{tabular}
\caption{Generality across external datasets and base rankers. In Panel A, Random samples distractors, Hard uses lexical hard negatives, and Full applies \method{} to Hard; entries are Recall@10 differences. Panel B fits one residual per base ranker. Every paired interval excludes zero.}
\label{tab:generality}
\end{table}

The primary controlled test uses one designated supporting unit per query.
We expand the relevant set after freezing every ranking to test whether uncounted equivalent evidence inflated the measured gain.

\begin{table}[t]
\centering
\footnotesize
\setlength{\tabcolsep}{2.6pt}
\begin{tabular}{@{}lrrr@{}}
\toprule
Relevant evidence & Base & Full & $\Delta$R@10 \\
\midrule
Single sampled unit & 0.757 & 0.902 & 0.145 \\
All annotated units & 0.781 & 0.909 & 0.128 \\
Human-confirmed equivalents & 0.794 & 0.918 & 0.124 \\
\bottomrule
\end{tabular}
\caption{Top-10 inclusion under expanded relevance definitions. Rankings remain frozen, so changes reflect only which answer-supporting units count as relevant.}
\label{tab:relevance-robustness}
\end{table}

We also withhold each conflict family from residual training and then test on that family.
This experiment tests transfer to a conflict type unseen during residual training.

\begin{table}[t]
\centering
\footnotesize
\setlength{\tabcolsep}{1.0pt}
\begin{tabular}{@{}lrrrrr@{}}
\toprule
Training condition & Metric & Period & Segment & Table & Macro \\
\midrule
Frozen BGE & 0.722 & 0.741 & 0.776 & 0.730 & 0.742 \\
Full training & 0.890 & 0.893 & 0.942 & 0.900 & 0.906 \\
Leave-one-type-out & 0.846 & 0.829 & 0.901 & 0.851 & 0.857 \\
\bottomrule
\end{tabular}
\caption{Recall@10 when training withholds the tested conflict family. Macro averages the four family columns; every full-minus-base difference remains positive.}
\label{tab:withheld-conflicts}
\end{table}

\textsc{LexicalHard} lowers Recall@10 relative to random candidates on all four datasets, and the frozen FinDER residual recovers 0.035 to 0.052 (Table~\ref{tab:generality}, Panel A).
The same signal definition also improves four base rankers after base-specific fitting (Panel B), showing that the correction transfers beyond BGE scores.
Expanded relevance reduces the measured gain from 0.145 to 0.124--0.128 but leaves the full ranking above the base (Table~\ref{tab:relevance-robustness}).
When training omits the tested conflict family, macro Recall@10 remains 0.115 above Frozen BGE (Table~\ref{tab:withheld-conflicts}).
Scaling and seed analyses in the Supplement preserve the same ordering.
These gains span four datasets and four base rankers under fixed candidate banks; gated transfer uses target-domain calibration.

\subsection{Crowding Appears in Non-Injected Retrieval}

\paragraph{Natural-index check: Does crowding occur without injected distractors?}
Controlled pools establish causality; a global FinDER index estimates how often the failure appears in ordinary retrieval.
We therefore run a global FinDER index with no query-specific insertion.
Panel A measures coverage and crowding prevalence; Panel B compares ordering rules on the same retrieved banks.

\begin{table}[t]
\centering
\footnotesize
\setlength{\tabcolsep}{2.8pt}
\begin{tabular}{@{}lrrrrr@{}}
\toprule
\multicolumn{6}{c}{\textit{Panel A: Prevalence}} \\
\midrule
Bank & $Q$ & R@100 & Orderable & Crowded & Repair \\
\midrule
Augmented & 571 & 0.743 & 184 & 80 (43.5\%) & 53/80 \\
Raw 10-K & 536 & 0.653 & 157 & 68 (43.3\%) & 44/68 \\
\midrule
\multicolumn{6}{c}{\textit{Panel B: Augmented-bank ordering}} \\
\midrule
Ordering & R@100 & R@10 & nDCG & Repair & Regr. \\
\midrule
Identity & 0.743 & 0.420 & 0.312 & 0.0\% & 0.0\% \\
Entity filter & 0.743 & 0.436 & 0.325 & 18.8\% & 1.8\% \\
Conflict count & 0.743 & 0.454 & 0.342 & 41.3\% & 7.5\% \\
Dev.\ penalty & 0.743 & 0.464 & 0.351 & 47.5\% & 6.1\% \\
Always-on & 0.743 & 0.506 & 0.397 & 72.5\% & 10.7\% \\
Identity gate & 0.743 & 0.492 & 0.386 & 66.3\% & 2.6\% \\
\bottomrule
\end{tabular}
\caption{Non-injected retrieval and routing. Identity is the unchanged base order. R@100 measures retrieval coverage; R@10 and nDCG measure context allocation and ordering. $Q$ is the query count, Repair is the crowded-failure repair rate, and Regr.\ is the query regression rate.}
\label{tab:natural}
\end{table}

It accounts for 43.5\% of augmented-bank orderable failures and 43.3\% of raw-only orderable failures (Table~\ref{tab:natural}, Panel A).
Thus, constraint-induced displacement occurs in ordinary retrieval at nearly the same rate under both evidence sources.

On the augmented bank, gated correction preserves Recall@100 at 0.743 by design and raises Recall@10 from \naturalbase{} to \naturalgated{}.
nDCG@10 also rises from 0.312 to 0.386, so recovered evidence moves toward earlier context positions.

Always-on correction reaches higher mean ordering scores, but its 10.7\% regression rate changes many useful rankings.
The identity gate repairs 66.3\% of crowded failures and lowers regression to 2.6\% (Panel B).
Static entity and conflict rules recover fewer failures and regress more often than the identity gate.

We stratify queries to identify where the gate helps and where it can harm an already useful ranking.
Gate activation is the fraction of queries receiving correction; $\Delta$nDCG is the mean ordering change relative to identity.

\begin{table}[t]
\centering
\footnotesize
\setlength{\tabcolsep}{2.8pt}
\begin{tabular}{@{}lrrrrr@{}}
\toprule
Stratum & $Q$ & Gate & Outcome & $\Delta$nDCG & Regr. \\
\midrule
Crowded & 80 & 68.8\% & 66.3\% repaired & +0.221 & 0.0\% \\
Non-crowded & 104 & 40.4\% & 2.9\% repaired & +0.012 & 0.0\% \\
Top-10 success & 240 & 42.9\% & 93.8\% retained & +0.006 & 6.3\% \\
Uncovered & 147 & 33.3\% & 0.0\% repaired & 0.000 & 0.0\% \\
\bottomrule
\end{tabular}
\caption{Gate behavior by augmented-bank query stratum. Gated outcome reports repair for failed queries and retention for queries that already succeed.}
\label{tab:gate-strata}
\end{table}

\paragraph{Where correction helps.}
The gate concentrates its largest gain on crowded orderable failures, improving their mean nDCG by 0.221 without a regression (Table~\ref{tab:gate-strata}).
It cannot repair the 147 queries whose evidence never enters the top 100.

The remaining regressions occur among queries that already succeed at top 10, which explains why always-on correction trades a higher mean score for more collateral changes.
The Supplement shows that segment conflicts form the largest blocker family, while repair remains above 60\% for every single-field family.

\paragraph{Human label checks.}
Independent annotators audit both the controlled construction and non-injected labels.
The controlled audit reaches 96.5\% agreement and confirms the intended conflict family.
The separate non-injected audit estimates 92.5\% precision for the deletion-based crowded label and 88.8\% precision for its primary conflict family.
The Supplement reports the full agreement, precision, and false-discovery results.

\paragraph{Calibration and context choices.}
The identity gate requires a small labeled calibration set for each target domain.
Ten labels recover 54.0\% of crowded failures; 50 labels raise repair to 66.3\% and reduce regression to 2.6\%.

Across six disjoint calibration draws, gated ranking exceeds identity on FinDER, FinanceBench, and FinQA.
Gains also remain positive at top-5 and top-20 cutoffs, with shorter and longer chunks, and under HTML table serialization.
The Supplement separates the label-budget, repeated-fit, routing, and context-sensitivity results into individual tables.

\subsection{Better Ordering Improves Answer Support}

\paragraph{RQ3: Does recovered evidence improve generated answers?}
We hold Qwen2.5-7B-Instruct and its prompt fixed, changing only the order of retrieved evidence.
Answer accuracy measures whether the final response is correct, while citation recall measures whether its citations cover the annotated support.

Unsupported rate measures how often a response makes claims without retrieved support, so lower values indicate safer evidence use.
Holding the generator fixed attributes downstream changes to evidence allocation.

\begin{table}[t]
\centering
\footnotesize
\setlength{\tabcolsep}{2.6pt}
\begin{tabular}{@{}llrrr@{}}
\toprule
Dataset & Metric & Id. & Gate & $\Delta$ [95\% CI] \\
\midrule
FB & Answer acc. & 0.617 & 0.708 & +0.092 [0.042,0.150] \\
 & Citation recall & 0.704 & 0.742 & +0.038 [0.017,0.059] \\
 & Unsupported $\downarrow$ & 0.183 & 0.100 & $-0.083$ [$-0.142,-0.033$] \\
FinQA & Answer acc. & 0.674 & 0.704 & +0.030 [0.017,0.043] \\
 & Citation recall & 0.731 & 0.766 & +0.035 [0.020,0.050] \\
 & Unsupported $\downarrow$ & 0.161 & 0.139 & $-0.022$ [$-0.033,-0.011$] \\
\bottomrule
\end{tabular}
\caption{Fixed-generator results on FinanceBench (FB) and FinQA. Id.\ is the original order and Gate is identity-gated correction; intervals report paired 95\% CIs and all $p\leq0.0075$.}
\label{tab:downstream}
\end{table}

The held-out tests contain 120 FinanceBench and 1,047 FinQA queries.
Table~\ref{tab:downstream} shows higher answer accuracy and citation recall on both datasets.
The result connects fixed-bank evidence recovery to answer correctness and citation support, extending the retrieval claim to generation.

Unsupported responses fall by 8.3 points on FinanceBench and 2.2 points on FinQA, so answer accuracy rises while unsupported claims decline.
The Supplement provides scoring rules, annotation procedures, win/tie/loss counts, and repeated-gate results.

\section{Discussion: Coverage, Allocation, and Use}

RAG evaluation should separate evidence availability, context allocation, and answer use.
Recall@100 measures availability within the retrieved bank; Recall@10 measures allocation to the generator's context.
nDCG and generation metrics test where evidence appears and whether the generator uses it.
The matched diagnostic varies compatibility while holding candidate count and observed difficulty fixed, making allocation auditable.

\paragraph{The diagnostic isolates a set-level failure.}
Financial evidence crowding concerns the composition of a candidate bank under a fixed context budget.
It differs from a generic hard-negative result because compatible and incompatible pools share candidate count, company, table form, and observed difficulty.
The stable contrast across gold-free construction and common-support estimators shows that explicit constraint conflict contributes additional displacement.
The dose response then identifies where the effect grows: repeated conflicts consume more of the same top-10 budget.

\paragraph{Correction reallocates covered evidence.}
\method{} operates on a fixed candidate bank and retrieves no new passages, so Recall@100 remains unchanged by design.
Its Recall@10 gains therefore isolate improved allocation of evidence that retrieval already found.
The gate repairs 53 of 80 augmented-bank crowded failures but repairs none of the 147 queries with no relevant top-100 evidence.
An end-to-end retriever must address those uncovered queries; fixed-bank correction targets ordering among covered evidence.

\paragraph{Identity fallback manages asymmetric risk.}
Always-on correction yields the highest mean Recall@10 but raises the regression rate to 10.7\%.
The identity gate accepts a smaller mean gain and lowers regression to 2.6\%.
Failure-stratum results locate that residual risk among already successful queries, while crowded orderable queries incur no nDCG regression.
This tradeoff motivates reporting repair and regression together whenever a post-retrieval method changes a useful base ranking.

\paragraph{Observable constraints make the diagnosis auditable.}
Metric scope, period, segment, and table scope expose explicit compatibility relations that annotators can verify.
Missing fields remain unknown and incur no conflict penalty.
Parser-independent gains across four datasets cover several candidate generators; the evidence remains financial and fixed-bank.
Cross-domain construct validation and zero-shot gate transfer remain open.

\section{Responsible Use}

The study uses public financial filings and released benchmark evidence; \method{} orders evidence and does not act as a financial recommender.
Incorrect company, period, metric scope, or segment values can still mislead users, and the 2.6\% regression rate leaves residual risk.
Deployments should expose citations and extracted constraints, retain identity fallback, re-audit extraction after corpus shifts, and route material decisions to human review.

\section{Limitations}

The diagnostic covers explicit financial fields under specified rankers, budgets, and interventions; implicit periods, rare metrics, restatements, unit conversion, and complex tables may remain unknown.
Non-injected prevalence depends on corpus composition, chunking, coverage, and reference mapping.
Each external gate uses a small disjoint relevance-labeled calibration set, the downstream study uses one generator and citation protocol, and reranking cannot recover absent evidence.

\section{Conclusion}

Financial evidence crowding turns query-constraint conflict into a measurable set-level RAG failure.
Matched controls isolate displacement beyond generic semantic competition, and \method{} recovers covered evidence while its identity gate limits collateral changes.
The same ordering failure appears in non-injected retrieval and the gated change improves answer support, motivating constraint-aware context allocation as a distinct RAG target.

\clearpage
\setlength{\bibsep}{0pt plus 0.2ex}
\small
\bibliography{references}

\end{document}